\documentclass[conference]{IEEEtran}
\usepackage{balance}
\usepackage{ifpdf}
\usepackage{cite}
\ifCLASSINFOpdf
   \usepackage[pdftex]{graphicx}
\else
\fi
\usepackage{amsmath}
\usepackage{algorithmic}

\usepackage{array}

\ifCLASSOPTIONcompsoc
  \usepackage[caption=false,font=normalsize,labelfont=sf,textfont=sf]{subfig}
\else
  \usepackage[caption=false,font=footnotesize]{subfig}
\fi

\usepackage{stfloats}
\usepackage{url}

\begin{document}
%
\title{Performance Dynamics and Success\\in Online Games}

\author{\IEEEauthorblockN{Anna Sapienza}
\IEEEauthorblockA{USC Information Sciences Institute\\
Marina del Rey, California, 90292}
\and
\IEEEauthorblockN{Hao Peng}
\IEEEauthorblockA{USC Information Sciences Institute\\
Marina del Rey, California, 90292}
\and
\IEEEauthorblockN{Emilio Ferrara}
\IEEEauthorblockA{USC Information Sciences Institute\\
Marina del Rey, California, 90292}}

\maketitle

\begin{abstract}
Online data provide a way to monitor how users behave in social systems like social networks and online games, and understand which features turn an ordinary individual into a successful one. Here, we propose to study individual performance and success in Multiplayer Online Battle Arena (MOBA) games. Our purpose is to identify those behaviors and playing styles that are characteristic of players with high skill level and that distinguish them from other players. To this aim, we study \textit{Defense of the ancient 2} (Dota 2), a popular MOBA game. Our findings highlight three main aspects to be successful in the game: \textit{(i)} players need to have a warm-up period to enhance their performance in the game; \textit{(ii)} having a long in-game experience does not necessarily translate in achieving better skills; but rather, \textit{(iii)} players that reach high skill levels differentiate from others because of their aggressive playing strategy, which implies to kill opponents more often than cooperating with teammates, and trying to give an early end to the match. 
\end{abstract}


%
\IEEEpeerreviewmaketitle

\section{Introduction}

The increasing availability of online data provides a key way to monitor and study how users behave in online social systems and to understand which are the characteristics that drive users to success, e.g., having a popular account on Twitter, being an experienced user on StackExchange (collecting badges), or becoming an advanced player in online games. A key research question is to understand how these behavioral patterns and users' performance are influenced: which are the major characteristics that bring users to perform better (reaching success)? Which characteristics have a bad influence on the user behavior (performance depletion)?

Individual performance and success were previously studied in different contexts: individual impact in scientific research~\cite{radicchi2009diffusion,sinatra2016quantifying}, successful learning strategies in education~\cite{romero2013data,rodi2015optimal}, online popularity~\cite{aral2012identifying,ferrara2014online,singer2016evidence}, sport performance~\cite{memmert2017current}, etc. Recently, online platforms have attracted much attention from our research community, as they provide a way to study new aspects related to human behavior. This is the case for social media like Twitter, whose data can be studied to identify the mechanisms behind user influence and popularity~\cite{cha2010measuring, hong2011predicting,aral2012identifying,kooti2016twitter}, Q$\&$A networks as StackExchange~\cite{movshovitz2013analysis,singer2016evidence,ferrara2017dynamics}, in which users acquire experience by collecting badges when fulfilling a specific task, and online games~\cite{huang2013mastering,bardzell2008blissfully,benefield2016virtual}, where players increase their skills by playing individually or by working as a part of a team. Here, we focus on Multiplayer Online Battle Arena (MOBA), a popular game genre in which users fight with their teammates to conquer the opponent base~\cite{ducheneaut2004social,ferrari2013generative}. In particular, we analyze a well-known MOBA game: \textit{Defense of the ancient 2} (Dota 2). This type of game provides indeed key insights on how users behave and improve their skills over time to reach success~\cite{drachen2014skill,yang2014identifying,pobiedina2013successful,pobiedina2013ranking}. 

By studying individual matches over time, we not only explore how players improve while playing consecutive matches, but also monitor how their performance changes based on the role they are impersonating~\cite{eggert2015classification}. In particular, we are interested in deepening our understanding of the way players either enhance or worsen their skills in the game. To this aim, we study actions of individual players in consecutive matches (i.e. sessions) to identify how and why their performance changes. We are indeed interested in addressing some of the following questions: how can we identify successful players? Are high skill levels only reached by long-time experienced players? What are the key features defining a successful player? What are the differences between high and low experience/skill players? 

Giving an answer to these questions could help to design new incentives (badges, rewards) with the aim of increasing player engagement in the game as well as providing useful recommendations to lower performer players to encourage them to change their attitude and achieve better results~\cite{deterding2011game}.

The present paper is organized as follows: in \S\ref{datamethods} we describe the data and the collection process used in the study; in \S\ref{meth} we introduce the methods used to perform the analysis; in \S\ref{results} we report the results of our study and compare low/high experience players and low/high skill players to understand which are the main characteristics of successful strategies in the game; in \S\ref{relwork} we comment the related work and provide a comparison with our results; finally, in \S\ref{concl} we draw the conclusions of our study, and describe the main findings and future work.

\section{Data description and collection}
\label{datamethods}

\subsection{Dota 2}

Dota 2 is one of the most popular Multiplayer Online Battle Arena (MOBA) video games providing a combination of action, strategy and role play. It is played by millions of both professional and casual fans every day worldwide.\footnote{\url{https://www.opendota.com/}}\footnote{\url{https://www.dotabuff.com/}} A single match involves a dynamic and fast-paced battle between two teams of 5 players competing to collectively destroy the base structure, i.e. the \textit{Ancient} defended by the opposing team. Each player in a match controls one of the 113 characters, i.e. \textit{heroes} available in the game. These heroes are an essential element of the game as having different combinations of heroes in the two opposing teams makes each match unique. 

Each hero is categorized, depending on its skills, in three separated groups, namely \textit{Strength, Agility} and \textit{Intelligence}, which in this paper we refer to as \textit{hero types}. Strength heroes are strong warriors that can survive receiving more damage than others and can withstand longer battles. Agility heroes are known for their quick response and attack speed. Intelligence heroes often utilize their magical powers to enhance the attack against others and can also support their allies, e.g., via healing or buffing skills, in different tasks. By design, a single hero cannot be perfect in all three dimensions, but the interplay between different types of heroes impacts the overall strategy and outcome of each match. On the battlefield, players accumulate experience and gold through a variety of actions (\textit{kill}, \textit{assist}, \textit{death}, etc.) that can be later spent to acquire new skills to improve one's hero and overtake the opponents.

Players usually play several matches consecutively before taking an extended break. On average, each match lasts for $40$ minutes. According to different ways of assembling players for a competition,\footnote{\url{https://dota2api.readthedocs.io/en/latest/responses.html#lobby-type}} all matches can be grouped into $8$ different types, namely Public matchmaking, Practice, Tournament, Tutorial, Co-op with AI, Team match, Solo queue, Ranked matchmaking, and Solo Mid $1$ vs $1$. In this paper, we only focus on matches with \textit{Public} or \textit{Ranked} matchmaking to ensure the presence of $10$ human players in each match. 

\subsection{Data Collection and Preprocessing}

We collected 3,566,804 matches spanning from July 17, 2013 to December 13, 2015 using the Dota 2 official API.\footnote{\url{http://dota2api.readthedocs.io/en/latest/}} Due to data sparsity, we perform our analysis on 3,300,146 matches played since May 2015. Each entry in our dataset contains match meta information, including start time, duration, lobby type (Ranked matchmaking, Public matchmaking, etc.) and winning status, and it also contains information about each player's performance in that match, such as number of kills, assists, deaths, gold earned, etc.

We preprocessed the collected data and selected all those matches for which we have access to complete information. In particular, we discarded all the matches characterized by one of the following cases:
\begin{itemize}
\item \textbf{Connection errors}: some matches end earlier due to connection errors. In this case, the winning status related to the match corresponds to a null value, and the actions performed by the players will be only partially recorded.
\item \textbf{Default players}: not all the players allow to make their personal information public. In this case, a unique identifier is assigned to all these players, thus preventing their identification.
\item \textbf{Leaver status}: some players quit a match at the beginning, thus leaving the two opposing team unbalanced. In this case, the player's feature fields (kills, assists, deaths, etc) will be $0$. 
\end{itemize}

After this filtering, we collect all  players that appear in the selected matches, which correspond to a total of 1,805,225 players.

\begin{figure}[tbh]
\centering
\includegraphics[width = 0.9\columnwidth]{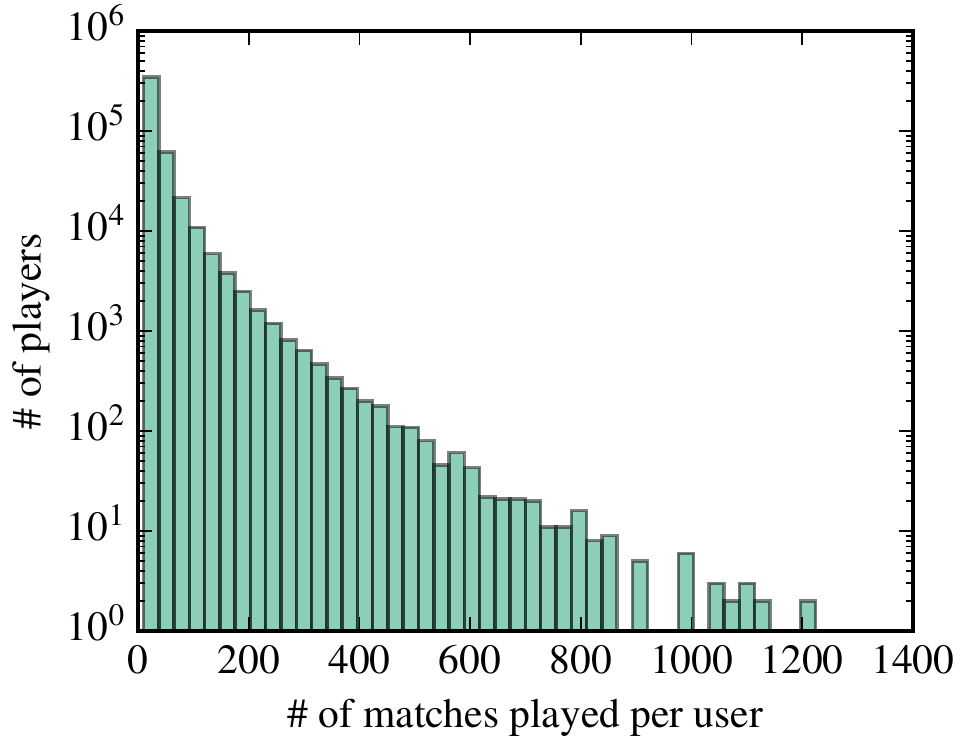}
\caption{Distribution of the number of matches played by each user over the total 1,805,225 players selected after the data preprocessing step.}
\label{nmatches}
\end{figure}

Finally, we compute the distribution of the number of matches per user among the 1,805,225 selected players. The distribution, reported in Fig.~\ref{nmatches}, shows that almost  $75\%$ of the players participated to less than $10$ matches. For our analysis, we require that each selected player played at least $10$ matches. The final number of player selected after applying this threshold is 460,026 over the 1.8 million initial ones.

\section{Methods}
\label{meth}

With the aim of investigating how player's performance evolves over time, we focus on two different aspects. First of all, we look at their performance in separated sessions: to this aim, we first grouped together matches that are played in sequence without extended breaks. Second, we are interested in understanding how players' performance changes accordingly to the type of hero they select and if there is any preference in the hero type selection. Thus, we study player's behaviors over time by dividing them in game sessions and further explore player's performance based on the selected hero by looking at sessions characterized by the same hero type. 

\subsection{Game Sessions}

\begin{figure}[h!]
\centering
\includegraphics[width = 0.9\columnwidth]{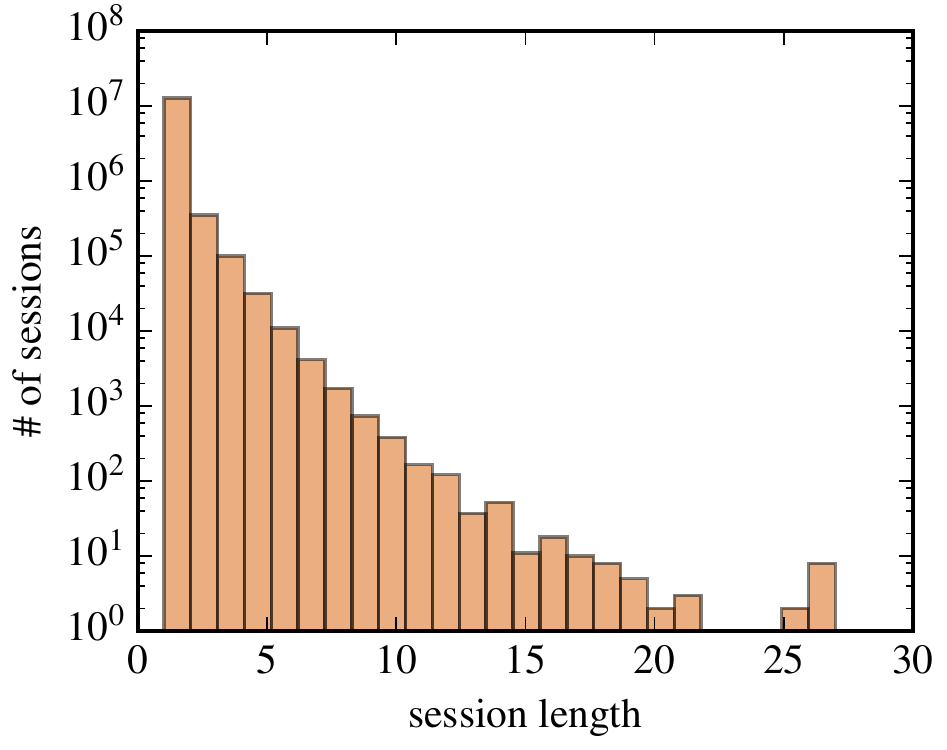}
\caption{Distribution of the number of sessions of different lengths.}
\label{nsessions}
\end{figure}

We divide our dataset in time series of matches played by those players having more than $10$ matches. The long sequence of each player's ordered historical matches can be divided into short time game sessions, namely short periods of playing behavior without an extended break. In this work, we identify game sessions by examining the time intervals between consecutive matches against a predefined threshold. We indeed define a game session to be  composed by several consecutive matches with no more than a $15$-minute break between them. This threshold corresponds to the peak of the distribution of the time breaks between all the matches in our dataset. By dividing matches into sessions, we expect that users' playing behavior, e.g. adopted strategies, may be different for matches in the same session. We report the distribution of the number of sessions with different lengths in Fig.~\ref{nsessions}. In the following, we will focus on the analysis of sessions whose length is between  $1$ and $5$ matches. This amounts for over 90\% of the total sessions.

\subsection{Game Sessions for Hero Types}

\begin{figure*}[t!]
\centering
\includegraphics[width = \textwidth]{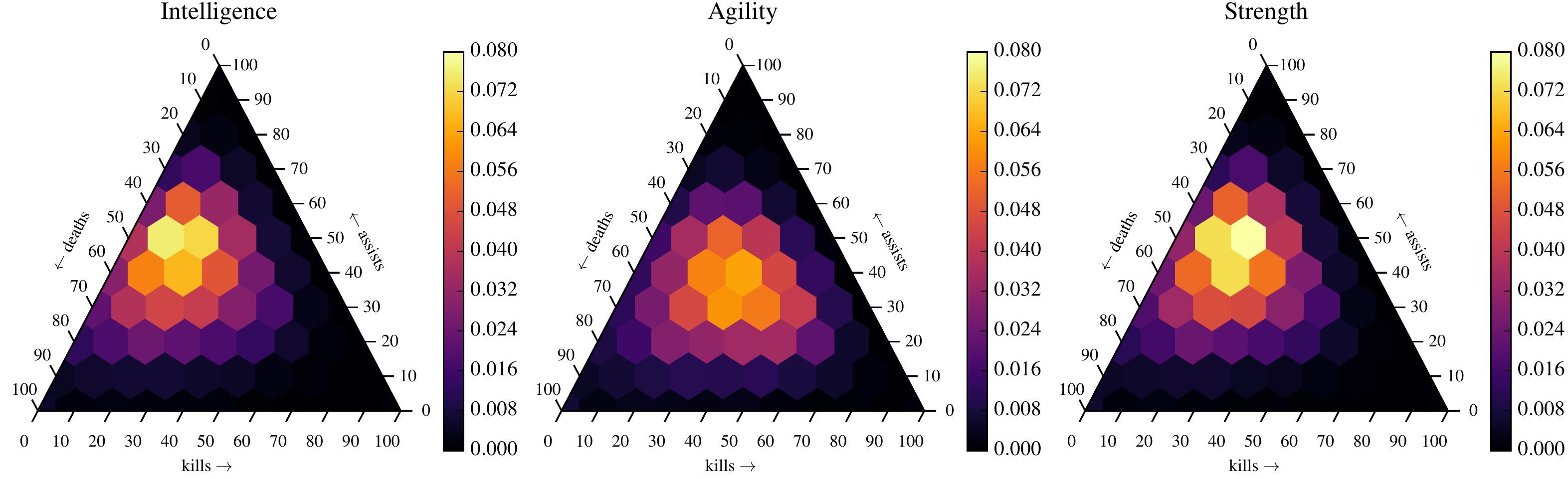}
\caption{Distribution of performed actions in a match divided by hero type. The three axes represent the percentages of kills, assists and deaths performed, while the color intensity represents the normalized number of matches for which a certain triplet is performed.}
\label{ternary}
\end{figure*}

\begin{table*}[t]
\centering
\caption{Number of sessions of length between  $1$ and $5$, divided by hero types.}
\label{sess-hero}
\resizebox{0.75\textwidth}{!}{
\begin{tabular}{c|c|c|c|c|c}
Hero type    & Sessions of length 1 & Length 2 & Length 3 & Length 4 & Length 5 \\ \hline
Intelligence & 4,310,247       & 288,211        & 34,375        & 5,860          & 1,233          \\ \hline
Agility      & 3,503,961       & 200,129        & 21,508        & 3,435          & 712           \\ \hline
Strength     & 3,375,104       & 184,659        & 18,845        & 2,920          & 577          
\end{tabular}
}
\end{table*}

A player must select a specific hero before starting a match. The decision is based upon compound factors: overall team formation and role balance, as well as player's preferences. Although different heroes have different skill sets and abilities, heroes of the same type often have similar strengths and weaknesses. Thus, with the expectation that a player controlling heroes of different types in a session may have different performance, we further identify game sessions in which a player selected the same hero type in all matches of a single session. We report the number of sessions divided by hero type in Tab.~\ref{sess-hero}. We can observe that in general there is the tendency to use more Intelligence heroes which have a tactic and supporting role, followed by Agility and Strength heroes.

\subsection{TrueSkill}

Players in Dota 2 are ranked by using a \textit{Matchmaking Rating} (MMR),\footnote{\url{https://dota2.gamepedia.com/Matchmaking_Rating}} whose value characterizes the skill level of a player. This value is used to generate the team in each game and it increases/decreases if a player wins/loses the game. 

In the available data, we do not have access to the information related to the official MMR of each player. For this reason, we choose to define the skill level of each player in our dataset by computing the player \textit{TrueSkill} after each match. This computation allows us to characterize players by different skill levels.
The TrueSkill is a rating system designed by Microsoft,\footnote{\url{https://www.microsoft.com/en-us/research/project/trueskill-ranking-system/}} whose aim is to rank players in online games according to their skill level. It was developed to be used for video game matchmaking on Xbox Live and it is based on the popular Elo rating system, used in professional chess.\footnote{\url{https://en.wikipedia.org/wiki/Elo_rating_system}} Contrary to the Elo rating system, the TrueSkill is specifically created for games with more than two players, like MOBA games, as opposed to 1 vs 1 games. 

This method represents a player's skill as normal distribution, characterized by two parameters: $\mu$, i.e. the average skill of a player, and $\sigma$, i.e. the level of uncertainty in the player's skill. To compute the TrueSkill of each player we rely on the open Python implementation of the \textit{trueskill} library,\footnote{\url{https://pypi.python.org/pypi/trueskill}} in which we use as starting values for each player the default values of the library: $\mu = 25$ and $\sigma = \frac{25}{3}$. Then, for each player, the results of all matches they played, as well as the TrueSkill scores of their opponents at the time of that match, are used to determine each player's current TrueSkill score. These scores are finally used to group users and identify high/low skill ones.

\subsection{Experienced players vs performers}

\begin{figure*}[t!]
\centering
\includegraphics[width = 0.8\textwidth]{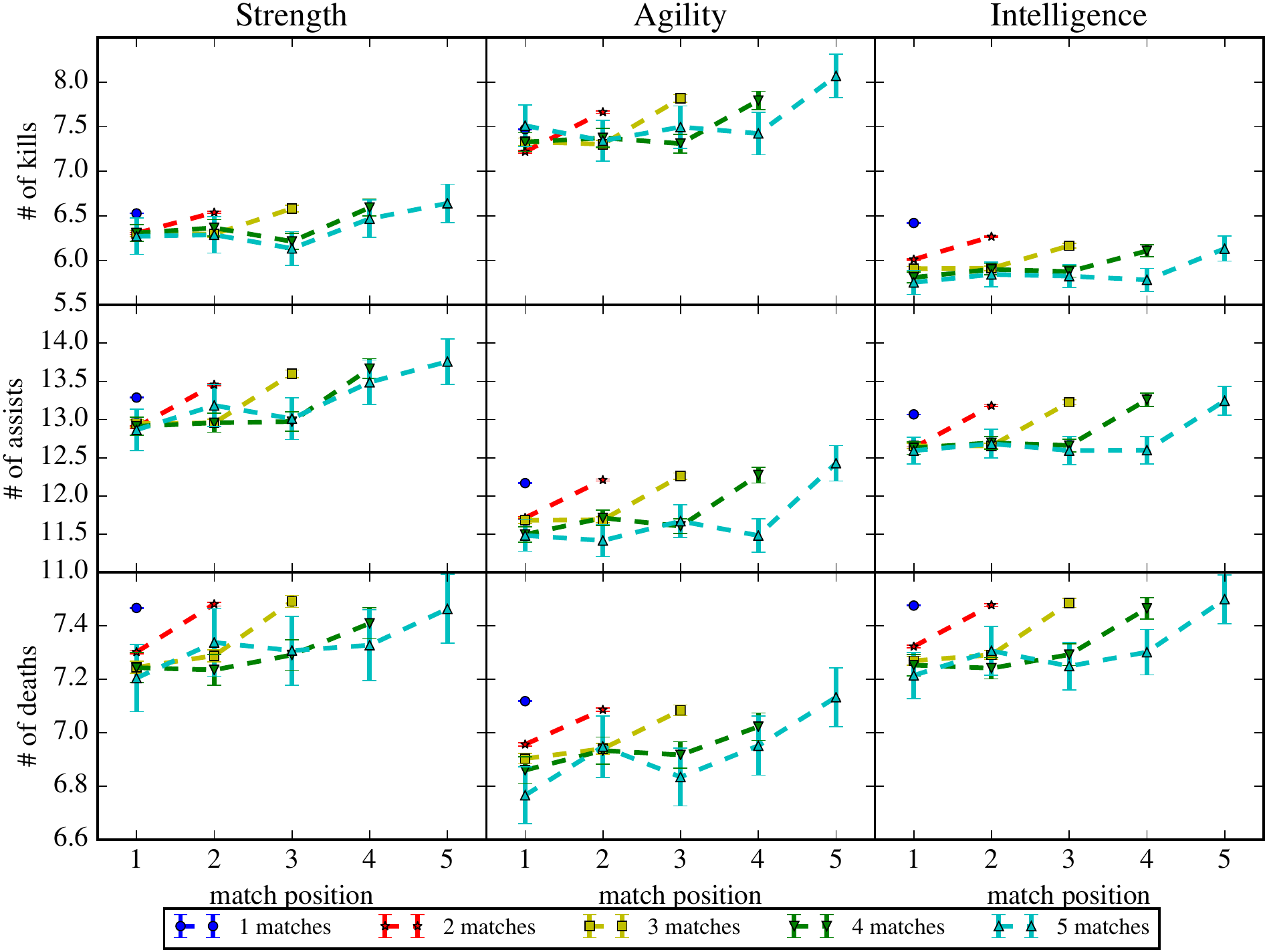}
\caption{Mean and standard deviation of the number of kills, assists and deaths along sessions of different lengths (from 1 to 5) divided by hero type.}
\label{kda}
\end{figure*}

With the aim of identifying the characteristics that make a player successful in the game, we divide players in two categories, i.e. high experience/skill players and low experience/skill players, according to two different metrics. In the first case, we divide the two groups of players by taking into account the total number of matches they played. We compute the distribution of the number of matches for all players having more than $10$ matches and we respectively label as \textit{low} and \textit{high experience players} those falling in the 5th and the 95th percentile of the distribution. 

In the second case, we divide the two groups by taking into account their experience in terms of skill level in the game. Thus, we select the TrueSkill of each player having more than $10$ matches, as defined in the previous section, and we compute the TrueSkill distribution. We respectively define, accordingly to the other group division, as \textit{low} and \textit{high skill players} those falling in the 5th and 95th percentile of the TrueSkill distribution. 
In the following section, we will report the results obtained by analyzing our dataset and we will compare the defined categorizations of players, to detect which characteristics make a player successful.

\section{Results}
\label{results}

\subsection{Hero type characterization}

We start our analysis by characterizing the three hero types, \textit{Intelligence}, \textit{Agility}, and \textit{Strength} and by studying how the players perform when using one of these heroes. We take into account both the hero type that is used by a certain player and the overall actions he/she performs in the match: number of kills, number of assists, and number of deaths. Fig.~\ref{ternary} shows the distribution of the number of actions in each match, in which we computed the percentages of kills, assists and deaths for each hero type. By comparing these ternary plots, we can characterize the different hero types, depending on the actions performed by players. The three distributions indeed highlight the different roles of the hero types. The Agility hero's distribution is more centered than the others. This is due to the fact that the kills performed by this hero type are more than in the other cases. This type of hero is indeed the one having the greatest offensive power. However, they also need support as they suffer a higher death rate in the first part of the game. On the contrary, the distributions related to both Intelligence and Strength heroes are similar. These distributions display higher number of assists and lower number of kills than the Agility heroes. This is due to the fact that Intelligence heroes usually have a support role, i.e., assist the team fight and keep allies alive. However, Strength heroes are characterized by lower death rates than Intelligence ones. They are indeed hard to kill and often start a team fight.  

\subsection{The warm-up effect}

\begin{figure*}[tbh]
\centering
\includegraphics[width = 0.8\textwidth]{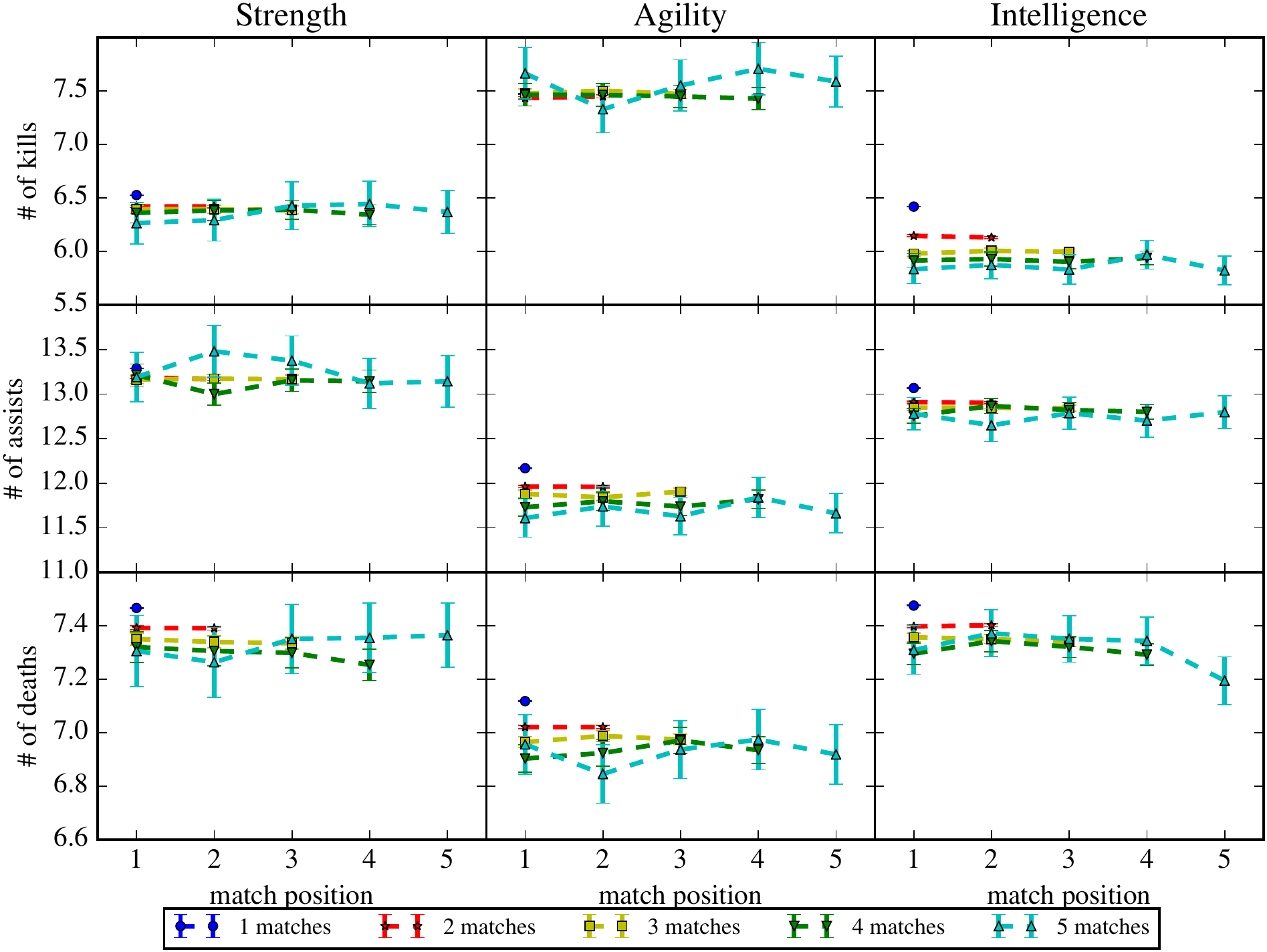}
\caption{Player's performance in terms of number of kills, assists, and deaths over randomized sessions of length from 1 to 5. Each session is given by the random shuffle of its matches.}
\label{rand-kda}
\end{figure*}

We are now interested in understanding how players perform over time, and if the use of a certain type of hero leads to better performance across consecutive matches. Thus, we divide player's histories in separate sessions, as defined in \S\ref{meth}, and we study sessions of different lengths: from 1 to 5. For each session of a given length, we compute the mean and the standard deviation of the number of kills, assists, and deaths, divided by hero type. 

The results are reported in Fig.~\ref{kda}, which provides a characterization of the hero type consistent with what we observed by looking at the ternary distributions over all the matches. On the one hand, players using Agility heroes perform a greater number of kills and lower number of assists and deaths if compared with the other two hero types. On the other hand, players using Strength and Intelligence hero types are more inclined in assisting during a team fight and sacrifice for the teammates (shown by the greater number of deaths). 

Moreover, these results show an interesting trend among the consecutive matches in a session: a \textit{warm-up effect}. During the same session indeed, the performance of players increases and this is particularly evident in the last match of each session. This rise in player's performance indicates that players warm-up along the matches and increment the final number of actions of almost  $10\%$. Another interesting feature is that the major increment occurs in the last match of the session. This could be interpreted by the fact that players, after the warm-up period, reach a satisfactory performance and thus decide to stop their playing session, i.e., ``leaving the game as winners.'' 

Finally, to verify the robustness of our results, we carry out our analysis on randomized sessions, which are computed by fixing the sessions of each player and shuffling at random the matches in the session~\cite{singer2016evidence,ferrara2017dynamics}. The results on the randomized data are shown in Fig.~\ref{rand-kda}, where the characteristic warm-up effect is not anymore present. Therefore, the flat trend across matches in the random sessions suggests that the improvement we observed during sessions is a significant result.

\subsection{Experience players vs performer players}

In the last part of our analysis, we aim at identifying the strategies and features that make a player successful. For this purpose, we study players from two different angles by comparing experienced players with skilled players. As introduced in \S\ref{meth}, we divide players into four groups: first we select low/high experienced players based on their seniority in terms of number of matches; then we select low/high skill players based on their TrueSkill scores, i.e. skill level. 

\begin{figure}[t!]
\centering
\includegraphics[width = 0.45\textwidth]{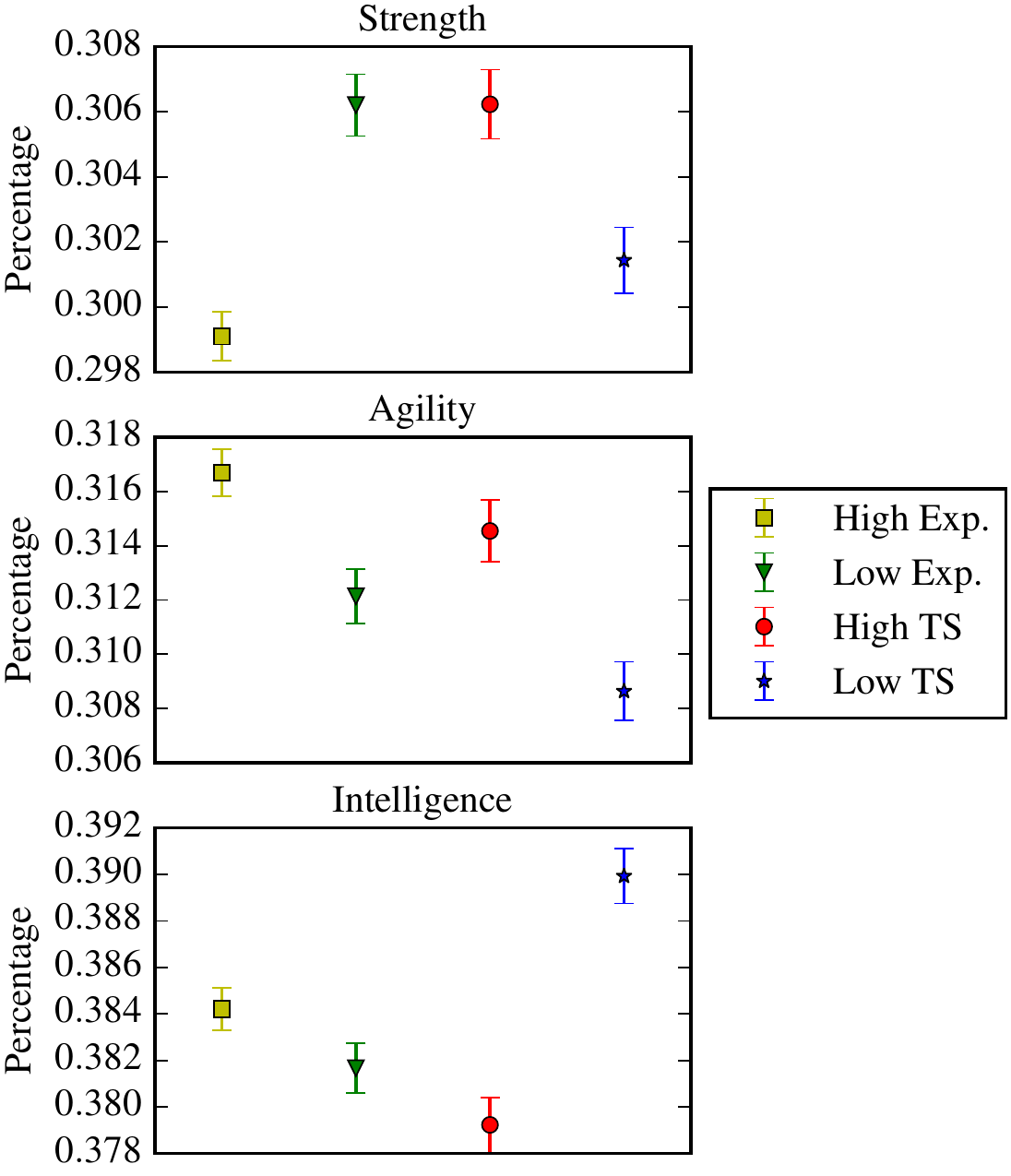}
\caption{Comparison between the percentage of hero types used by low/high experience players and low/high TrueSkill (TS) players. Each category, divided in the two groups (low/high), shows different preferences in the choice of the hero type.}
\label{prefhero}
\end{figure}

We compare these two groups of players (experienced vs skilled) by looking at their preferences when selecting an hero type. We compute the percentage of hero types selected by each player in the different categories and report in Fig.~\ref{prefhero} the corresponding mean and  standard error. The figure shows that the more experience a player has, the more he/she tends to prefer Intelligence and Agility roles, and thus both supporter and fighter heroes. However, this scenario is different when looking at the skill level of players. Players with high skill level play less supporter roles in proportion and prefer fighter hero types such as Agility and Strength heroes. This result also suggests that there is no direct correlation between successful (high skill) players and longtime (experienced) players.

This discrepancy between experienced players and skilled players is also reflected in their strategies and actions during matches. In the first case, both low and high experience players show a distribution of performed actions which is consistent with the one of Fig.~\ref{ternary} computed over all the players in the dataset. There is no evident difference in the collective strategies of low and high experience players, thus supporting the conclusion that playing longer does not necessarily lead to be successful in the game.

\begin{figure*}[tbh]
    \centering
  \subfloat[Low TrueSkill players]{
       \includegraphics[width=0.8\linewidth]{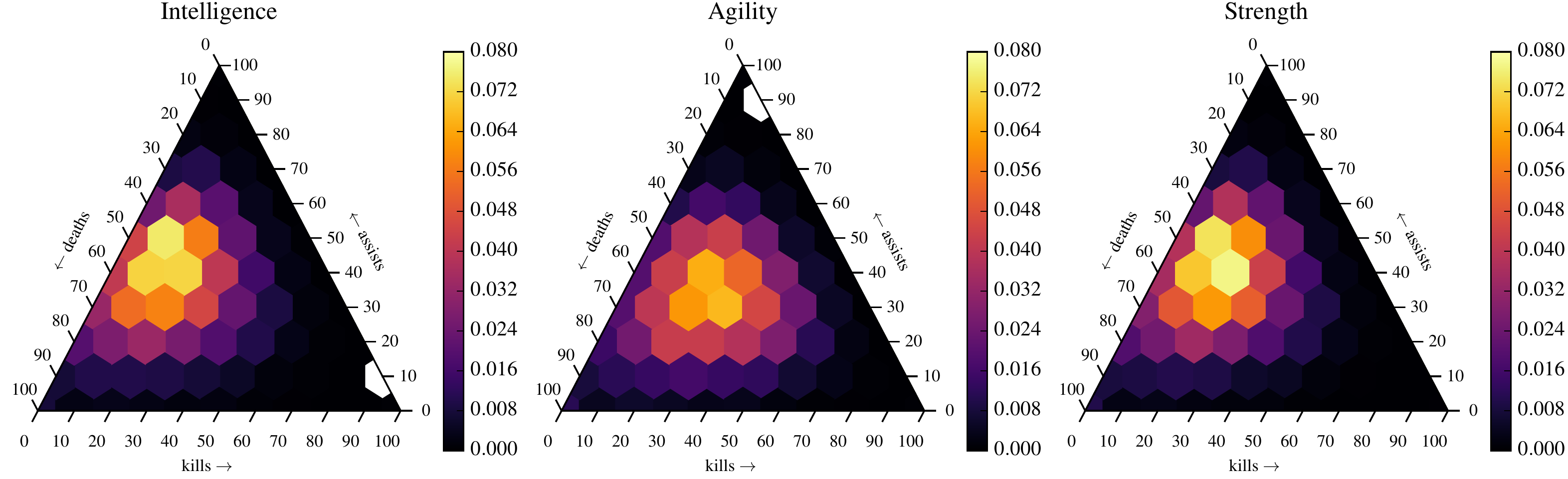}}
    \label{2a}\hfill
  \subfloat[High TrueSkill players]{
        \includegraphics[width=0.8\linewidth]{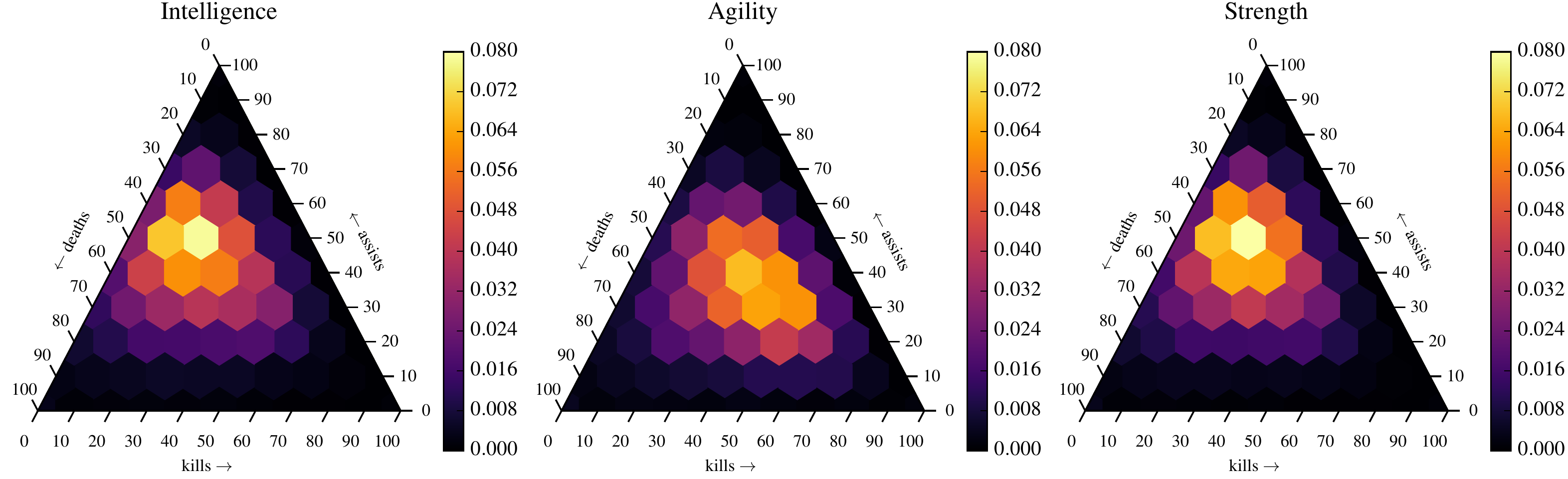}}
    \label{2b}
  \caption{Distribution of performed actions in a match divided by hero type and player's skill level. The three axes represent the percentages of kills, assists and deaths, while the color intensity represents the normalized number of matches for which a certain triplet is performed.}
  \label{ternTS} 
\end{figure*}

On the contrary, the comparison between performed actions during the matches played by low and high skill players highlights their discrepancies and characteristic features. As reported in Fig.~\ref{ternTS}a, low skill players are characterized by distributions that are skewed to the left side of the ternary plots, thus indicating that this group of players generally performs a low number of kills while prefers to assists teammates in combat. High skill players are instead characterized by distributions that are more centered in the ternary plots of Intelligence and Strength heroes, while in the case of Agility heroes the distribution tends to be skewed on the right part of the plot (\textit{cf.} Fig.~\ref{ternTS}b). These features indicate that in general high skill players make more kills than other players, even in the case in which their role is to support the teammates. Moreover, when using Intelligence or Strength heroes they sacrifice (die) themselves less than low skill players. 

Finally, we investigate if the two categorizations of players (experience/skill) provide any other insight that could help in characterizing a successful player. We look at the total duration of the matches played by low/high experience players and low/high skill players, to highlight groups differences. 

\begin{table}[h]
\centering
\caption{Mean and standard error of the match durations divided by low/high experience players and low/high TrueSkill players.}
\label{dur}
\resizebox{0.35\textwidth}{!}{
\begin{tabular}{c|c|c}
Player group      & mean duration & SE   \\ \hline
Low exp. players  & 2,532 s        & $\pm 0.34$ s\\ \hline
High exp. players & 2,507 s        & $\pm 1.16$ s\\ \hline
Low TS players    & 2,532 s        & $\pm 0.87$ s\\ \hline
High TS players   & 2,447 s        & $\pm 1.05$ s
\end{tabular}
}
\end{table}

We computed  mean and  standard error over the four groups of players, which are reported in Tab.~\ref{dur}. According to  previous results on low/high experience players strategies, the difference displayed in the average duration of the matches between these two groups is  marginal ($\simeq$25s). On the contrary, high skill players display a bigger difference if compared with low skill players ($\simeq$85s). In particular, players with higher skill level play shorter matches than those with low skill level, whose average duration is closer to the one related to low/high experienced players. A possible explanation for this phenomenon could rely on the more aggressive strategy that high skill players use. They indeed tend to perform more kills than others, even when they are interpreting support roles, and this playing style likely causes an early end to the matches.

In conclusion, our findings delineate as successful players in Dota 2 the more individualistic players, i.e., those that prevail over others by killing more even when they should interpret a support role and try to end the match as soon as possible by conquering the opponent base thanks to power plays.

\section{Related Work}
\label{relwork}

A great amount of work has been devoted to analyze several facets in MOBA games to identify common behavioral patterns, their evolution as well as characteristics that bring a team in a match to be successful and win the game. 

Drachen and collaborators~\cite{drachen2014skill} explored how game behaviors change and affect team skill levels. They analyzed the spatio-temporal behaviors of teams during a match. They looked at the temporal distribution of the distances between players in the same team and how this distribution varies between high/low skill teams. Eggert \textit{et al.}~\cite{eggert2015classification} were interested  in classifying player's behaviors via machine learning. These studies are focused on quantifying collective behaviors of teams in Dota 2 and relate these behaviors to player's roles that are not the official hero types we used in our analysis. 

Other techniques were developed to detect and rank the features in a team fight that help to predict whether a team will win a match or not~\cite{yang2014identifying,pobiedina2013successful,pobiedina2013ranking,sapienza2017non}.  In these studies, authors analyzed snapshots of a combat and described the action as a network, to extract common patterns leading to winning the match; the authors ranked factors that can be extrinsic to the game itself, by studying social ties of players in a team. 

We differentiate from these research directions as in the present paper we are mainly interested in providing an individual characterizations of players through both the performed actions and the interpreted role. Moreover, we define a player to be successful if his/her skill level (TrueSkill) is high, while in the existing literature the overall team level is analyzed and defined by the collective rate of victories/losses. 

\section{Conclusion}
\label{concl}

Online platforms, such as social networks, Q$\&$A sites, and  online games, provide a wealth of information, whose details can help identifying how people behave in different contexts. In particular, they are a useful means to detect the major characteristics that turn an ordinary user in a successful one. 

Here, we analyzed multiplayer online battle arena (MOBA) games, and in particular Dota 2, with the main purpose of extracting the underlying characteristics that distinguish players with a low skill level from those that in the course of their gaming experience reached the higher skill level. We decided to focus on one important aspect of the game: the characters players impersonate, namely heroes. By following the official hero categorization of Dota 2, we divided the $113$ heroes of the game in three main types: Intelligence, Agility, and Strength. These three hero types define a character on the basis of their ability and role, from support roles (as heroes in the Intelligence group) to fighter roles (as Agility heroes). 

As a first step in our evaluation, we characterized these three hero types by looking at the way in which they are used by Dota 2 players. We observed that overall, Intelligence heroes are used more than the other two types, probably because this supporting role is seen as more central to the game dynamics, providing more flexible heroes that can be more enticing to play. We also showed that players use heroes according to their ability and strength: Intelligence heroes to assist, Strength heroes to start team fights as well as assisting teammates, and Agility heroes to kill enemies. 

One interesting result we discussed is that players need some time to warm up and increase their performance across matches. By inspecting players' performance in consecutive matches, we found that the last match in a session corresponds to the one having higher performance in terms of number of actions (kills, assists, deaths). The robustness of this result is proved by randomizing the matches within sessions: the resulting null model does not display a significant ascending/descending trend --- on the contrary it remains constant from one match to the next --- corroborating our finding. 

Finally, we divided players by their experience (total number of matches played) in the game and by their skill level, by computing the so called TrueSkill. Within these two groups, we identified low/high experience players and low/high skill players. In-depth comparisons led to the conclusion that playing longer and thus having higher experience does not necessarily imply that one's skill level will increase. This finding is supported by the analysis of the distributions of actions performed by players with different hero types. Low/high experience players exhibit distributions which are consistent with those computed over all the players in the dataset, and thus do not show a specific playing style. On the contrary, the investigation of low/high skill players led to the characterization of  successful play styles. Successful players, which here correspond to those having high skill level (i.e., high TrueSkill), are aggressive players who prefer to kill enemies more than assisting teammates in a fight even if this is the role they are interpreting. This behavior leads these players to prefer the use of Agility and Strength hero types and to try to trigger an early end in the game with power plays. 

Future work will be devoted to further inspect the characteristics that make players successful. We will look for possible differences in the playing style of high skill players between Ranked and Public matches. This will help us understand if being part of a team of players of different levels could be an incentive for low skill players to perform better by learning from high skill players. Another possible direction will be to investigate how the TrueSkill of players evolves in time and if the playing styles of successful players we uncovered in the present work are constant or learned over time.

\section*{Acknowledgment}
The authors are grateful to DARPA for support (grant \#D16AP00115). This project does not necessarily reflect the position/policy of the Government; no official endorsement should be inferred. Approved for public release; unlimited distribution.

\balance
\bibliographystyle{IEEEtran}
\bibliography{biblio}

\end{document}